\documentclass[lettersize,journal]{IEEEtran}
\usepackage{amsmath,amsfonts}
\usepackage{algorithmic}
\usepackage{algorithm}
\usepackage{array}
\usepackage[caption=false,font=normalsize,labelfont=sf,textfont=sf]{subfig}
\usepackage{textcomp}
\usepackage{stfloats}
\usepackage{url}
\usepackage{verbatim}
\usepackage{graphicx}
\usepackage{cite}
\usepackage{balance}

\begin{document}

\title{Low-density parity-check codes: comparing \\cluster graph to factor graph representations}

\author{J du Toit, J du Preez, R Wolhuter
\thanks{J du Toit is with the Department of Electrical and Electronic Engineering, Stellenbosch University, Stellenbosch, South Africa (e-mail: jacowp357@gmail.com)}
\thanks{J du Preez is with the Department of Electrical and Electronic Engineering, Stellenbosch University, Stellenbosch, South Africa (e-mail: dupreez@sun.ac.za)}
\thanks{R Wolhuter is with the Department of Electrical and Electronic Engineering, Stellenbosch University, Stellenbosch, South Africa (e-mail: wolhuter@sun.ac.za)}
\thanks{Manuscript received April, 2023; revised April, 2023.}
}



\maketitle

\begin{abstract}
In this letter, we propose a novel inference technique for LDPC codes based on cluster graphs. 
Cluster graphs have superior inference characteristics over commonly used factor graphs. 
We also introduce a layered message passing schedule for regular and irregular cluster graph LDPC codes. Simulation results show that cluster graph LDPC codes outperform factor graph LDPC codes in terms of error correction and require fewer iterations to converge. Both graphical representations are evaluated using the sum-product and max-product message passing algorithms with BPSK signals transmitted over the AWGN channel. Our proposed approach provides a promising alternative to infeasible fully tree-structured LDPC codes.
\end{abstract}


\section{Introduction}
\label{sec:introduction}
Low-density parity-check (LDPC) codes represented by factor graphs usually have cycles, requiring an iterative loopy belief propagation (LBP) algorithm for inference, without guaranteed convergence or exact solutions. Conversely, message passing on cycle-free graphs produces exact marginals~\cite{koller2009probabilistic}. ``Clustering'' or ``stretching'' variable and factor nodes is a known concept in the channel coding community to remove cycles, but increases computational complexity due to larger domain sizes of the resultant clusters~\cite{kschischang2001factor}. Essentially, the ``clustering'' and ``stretching'' techniques discussed in \cite{kschischang2001factor} describe a similar process as that of constructing \emph{junction trees}. These techniques have been applied in detector architectures, but not on LDPC codes directly~\cite{colavolpe2022multiuser,colavolpe2005application,colavolpe2006ldpc}.

We consider a more general probabilistic graphical model (PGM) framework called \emph{cluster graphs} as formally defined in~\cite[Chapters 10 and 11]{koller2009probabilistic}. Although its name may suggest so, a ``cluster graph'' is not a representation that primarily attempts to group variables or factors into larger sets of variables or factors such as the ``clustering'' and ``stretching'' techniques discussed in~\cite{kschischang2001factor} (although it certainly can support that as well). A cluster graph typically preserves the original factors exactly as they are. 

While cluster graphs are not necessarily tree-structured, they must satisfy specific structural constraints, which is a step towards a fully tree-structured PGM. The \emph{layered trees running intersection property} (LTRIP) algorithm, developed in~\cite{streicher2017graph,streicher2021strengthening}, automatically compiles cluster graphs from an input set of factors. Cluster graphs offer a unique advantage over factor graphs by allowing messages passed between clusters to involve joint distributions. This preserves correlations between variables that result in superior inference characteristics, including faster convergence speeds~\cite{streicher2017graph,streicher2021strengthening}. 

We introduce cluster graphs as a novel approach to LDPC channel coding and evaluate its performance against traditionally used factor graphs. Cluster graphs do not follow the same message passing order as factor graphs and so we also present an approach to schedule message passing for both regular and irregular cluster graph LDPC codes.

The remainder of this letter is organised as follows. Section~\ref{sec:representation} explains how LDPC codes are represented as cluster graphs. The message passing approach and schedule are provided in Section \ref{sec:message-passing}, followed by our results in Section \ref{sec:results}. Finally, we conclude and discuss future work in Section~\ref{sec:conclusion}.

\section{Representation of LDPC codes}
\label{sec:representation}
\begin{figure*}%
    \centering
    \includegraphics[scale=0.38]{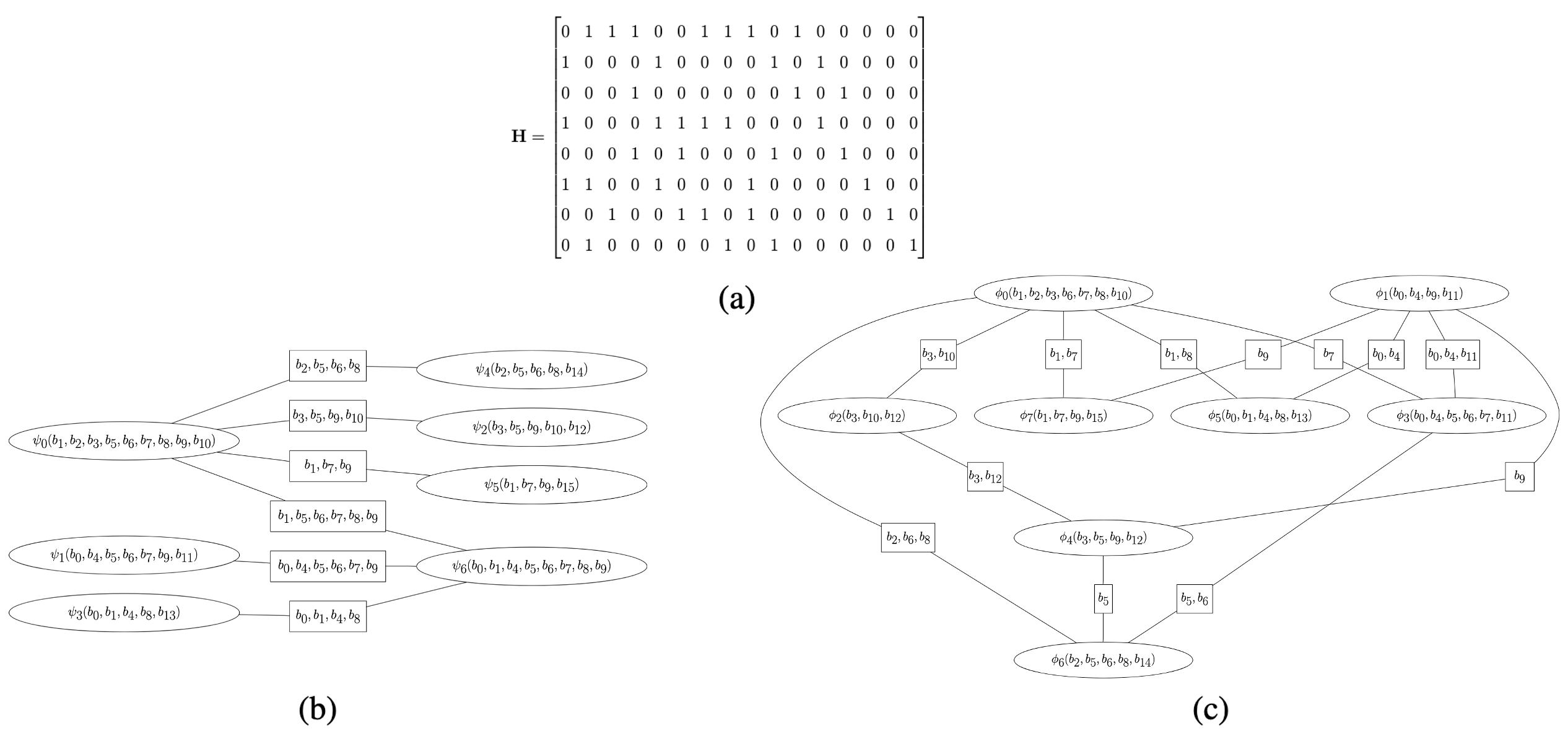}
    \caption{(a) $\mathbf{H}$ matrix of an irregular (16,8) LDPC code, with
(b) a junction tree representation, and (c) a cluster graph
representation compiled using the LTRIP algorithm. Note the increased cluster scopes in the junction tree of (b) as compared to the cluster graph of (c).}
    \label{fig:ldpc-graphs}
\end{figure*}

We use a small irregular (16,8) LDPC code to illustrate the differences between a factor graph, junction tree, and cluster graph representation. The $\mathbf{H}$ matrix is shown in Fig.~\ref{fig:ldpc-graphs}(a). We denote the codeword bits as $b_0$, ..., $b_{15}$ and the parity check factors as $\phi_0$ to $\phi_7$.

In a factor graph (or Tanner graph), nodes are connected to corresponding variable nodes whenever a factor depends on that particular variable. We consider a more general probabilistic framework, called \emph{cluster} graphs in which a factor graph representation is a special case (also known as a Beth\'e graph). Cluster graphs have two types of nodes, a \emph{cluster} node (ellipse) is a set of random variables, and a \emph{sepset} (short for ``separation set'') node (square) is a set of random variables shared between a pair of adjacent clusters. The sepset dictates the scope of the message passed between two clusters. Each parity check cluster contains typically one (but could be more) factor with all of its variables present in the scope of the cluster.

Using the junction tree algorithm, variables are clustered using the same ideas as presented in \cite{kschischang2001factor} to produce a cycle-free LDPC code as demonstrated in (b). While exact inference is possible on this small graph, the junction tree algorithm is NP-complete for large LDPC codes and the resulting clusters (e.g., $\psi_0$ and $\psi_6$) have increased scopes.

A cluster graph must satisfy specific structural constraints that impose a tree-like structure on each unique variable in the graph. This is referred to as the \emph{running intersection property} (RIP)~\cite{koller2009probabilistic}. A path between clusters requires clusters and sepsets to share one or more variables. Any pair of clusters sharing a common variable must be linked via a unique path (i.e. without loops) for that particular variable. The cluster graph in (c) is constructed using the \emph{layered trees running intersection property} (LTRIP) algorithm~\cite{streicher2017graph}. A factor graph also satisfies the RIP due to its star-like topology but contains univariate sepsets. It should be noted that the parity check constraints in (a) correspond to the clusters in (c). 

The LTRIP algorithm can be applied to large LDPC codes and proceeds in layers by considering each variable in a separate layer. For each variable, it determines a tree structure over all clusters. The sepsets between pairs of clusters are then merged across all layers to form the final sepsets. The resulting cluster graph satisfies the RIP and allows richer information content to be shared between clusters. The resultant graph structure in general will contain loops, i.e. it is not necessarily a tree structure. We refer the reader to \cite{streicher2017graph} for more detail regarding the LTRIP algorithm.

In contrast to what is the case with a factor graph, the sepsets in a cluster graph may in general contain more than one variable. The advantages of larger sepsets are that it contributes to faster convergence and less computation since fewer variables are marginalised out during message passing. The largest cluster nodes in the graph will dominate the computational requirements, and they are exactly the same as the largest factor nodes in the factor graph.

\section{Message passing approach}
\label{sec:message-passing}
We implement both sum-product (SP) and max-product (MP) inference. Note that we use MP as described in \cite[Section 13.4]{koller2009probabilistic} where sum is replaced with max (i.e., max-normalisation and max-marginalisation) to obtain the maximum posterior solution over all possible codewords using message passing. Our study uses a variant of LBP called \emph{loopy belief update} (LBU), also known as the Lauritzen-Spiegelhalter algorithm~\cite{lauritzen1988local}. The main differences between LBP and LBU are:
\begin{itemize}
    \item LBU uses cluster beliefs and sepset beliefs to express message passing,
    \item with LBU, cluster beliefs are updated with messages from all its connections and can be more informative compared to LBP,
    \item with LBU, sepsets are used to update a target cluster belief, and require only one sepset belief division. 
\end{itemize}

\subsection{Message passing schedule}
In LDPC codes represented as factor graphs, the flow of messages typically follows a path from the variable nodes to the factor nodes, and then back from the factor nodes to the variable nodes. The edges between clusters in our cluster graph are determined by the LTRIP algorithm and require a different message passing schedule. We introduce a layered cluster message passing schedule which aims to reduce computation further by isolating the largest parity check clusters in the graph. We use the same LDPC code from Section~\ref{sec:representation} to explain our message schedule.
\label{sec:message-passing-schedule}
\begin{figure*}[t]
\centering
\includegraphics[width=\textwidth]{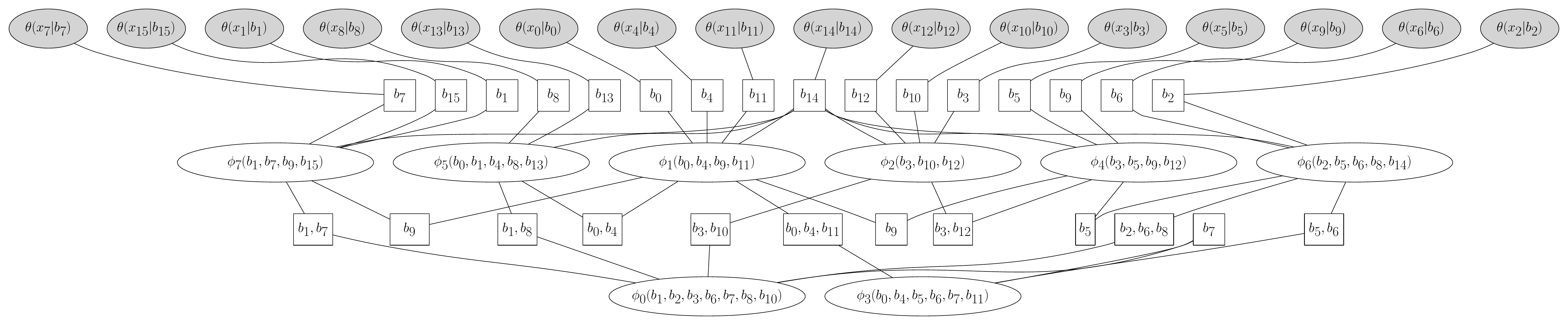}
\caption{A cluster graph (16,8) LDPC code with larger clusters at the bottom and conditional Gaussian clusters linked to the smaller parity check clusters.}
\label{fig:ldpc16}  
\end{figure*}

A message passing schedule is an important consideration for loopy graphs since the message order can influence convergence speed, accuracy, and the computational cost of inference. In loopy graphs, information can propagate from a cluster and continue along a path that eventually ends at the same cluster without traversing the same edge twice. Although not empirically verified here, these feedback loops (or cycles) may reinforce inaccurate cluster beliefs causing self-fulfilling belief updates, which affect the LDPC decoder's performance. This problem is more prominent in LDPC codes with small feedback loops as described in~\cite{ohtsuki2007ldpc}. Taking this into consideration, our message passing schedule: (1) uses a structured schedule with a fixed computational cost, (2) aims to minimise the effect of loops, and (3) aims to minimise the computational cost of inference. Message passing should proceed in layers so that the result of previous messages can be used immediately for following layers as message passing proceeds.

The message passing schedule is determined by first identifying the larger parity check clusters in the graph. We select the larger clusters $\phi_0$ (with cardinality 7) and $\phi_3$ (with cardinality 6). The message schedule starts with the selected clusters as initial sources $\mathcal{S}$ and proceeds by visiting all its neighbouring clusters $\mathcal{N}$, which become the immediate next layer of clusters. A set of available clusters $\mathcal{A}$ is kept to make sure that clusters from previous layers are not revisited, which helps minimise the effect of loops. We repeat this procedure to add subsequent layers of clusters until all clusters are included. The source-destination pairs are stored in a message schedule $\mathcal{M}$. This procedure isolates the initially selected large parity check clusters from the rest of the clusters as shown in Fig.~\ref{fig:ldpc16}. The idea is to keep the expensive clusters at the final layer so that the smaller (less expensive) parity clusters, in preceding layers, can resolve most of the uncertainty about the even parity states. When the larger parity clusters get updated, some of the even parity states in their discrete tables may have zero probability, which are removed due to our software implementation. This further reduces a large parity cluster's computational footprint. Our layered message passing schedule is detailed in Algorithm 1.

We use an additive white Gaussian noise (AWGN) channel, where $x_i$ are observed signal values and $b_i$ the bit probabilities. The observed conditional Gaussian clusters $\theta$ (also referred to as the intrinsic probabilities from the channel) are coupled to the parity check clusters in the layer furthest away from the initial isolated group of large clusters -- starting with the smallest parity clusters. This avoids expensive computational cost between the observed random variables and the larger parity clusters (even though we only need to multiply them in once). The isolated parity check clusters make up the bottom layer. Note that we do not link conditional Gaussian clusters to parity clusters in intermediate layers. We avoid this so that evidence enters the graph from one end (the top) and updates the latent clusters one layer at a time. Only if the first layer of parity check clusters do not have all the unique bits, the following layers are utilised until all conditional Gaussian clusters are connected.

\begin{algorithm}
\caption{Layered message passing schedule}\label{alg:alg1}
\begin{algorithmic}[1]
\STATE $\mathcal{S} \gets$ set initialised to large cluster IDs 
\STATE $\mathcal{A} \gets$ set initialised to all cluster IDs 
\STATE $\mathcal{M} \gets$ empty vector of pairs
\WHILE{$\mathcal{A}$ is not empty}
\STATE $nextLayer \gets$ empty set of integers
\FOR{$s$ in $\mathcal{S}$}
\STATE $\mathcal{A}$.erase($s$)
\STATE $\mathcal{N} \gets$ all neighbours of $s$
\FOR{$n$ in $\mathcal{N}$}
\IF{$n \in \mathcal{A}$}
\STATE $\mathcal{M}$.push\_back(pair($s$, $n$))
\IF{$n \not\in \mathcal{S}$}
\STATE $nextLayer$.insert($n$)
\ENDIF
\ENDIF
\ENDFOR
\ENDFOR
\STATE $\mathcal{S} \gets nextLayer$
\ENDWHILE
\RETURN $\mathcal{M}$
\end{algorithmic}
\label{alg1}
\end{algorithm}

Once the observed conditional Gaussian clusters updated the first parity cluster layer they are not needed further during message passing. Message passing continues towards the final parity cluster layer, which we refer to as the forward sweep. The backward sweep returns in the opposite direction, which concludes one iteration of message passing.

The following settings and software implementations apply to our inference approach:
\begin{itemize}
    \item a cluster is deactivated during message passing when messages entering it have not changed significantly from what it was in the previous round. This is determined by a symmetrical\footnote{The average divergence from both directions is used, taking into consideration the mode-seeking and moment-seeking behaviours of the metric given by $\frac{D_{KL}(P||Q) + D_{KL}(Q||P)}{2}$.} Kullback-Leibler divergence measure between the newest and immediately preceding sepset beliefs.
    \item the stopping criterion for inference is when a valid codeword was detected (also known as a syndrome check) after all parity check clusters ``agree'' on their shared bit values\footnote{Note that this is an additional requirement as compared to the standard factor graph implementation.} or when a maximum number of iterations is reached,
    \item all discrete table factors support sparse representations to reduce memory resources,
    \item zero probability states in discrete tables are removed during inference. 
    \item when using MP inference, for each parity check cluster a small inconsequential quantity of random noise is added to the even parity values to ensure a unique maximum assignment for each cluster.
\end{itemize}

For regular LDPC codes, all parity check clusters have the same cardinality. Our message passing schedule can still be used for such codes by selecting any one of the parity check clusters as the initial source cluster. We emphasise that our message passing schedule may not be the most effective message ordering for cluster graphs in general or its application to LDPC codes. This particular schedule was useful in another of our studies~\cite{jacojohancluster2}, which required additional prior distributions for channel noise estimation. 

\section{Results}
\label{sec:results}
\begin{figure*}[t]
\centering
\includegraphics[scale=0.34]{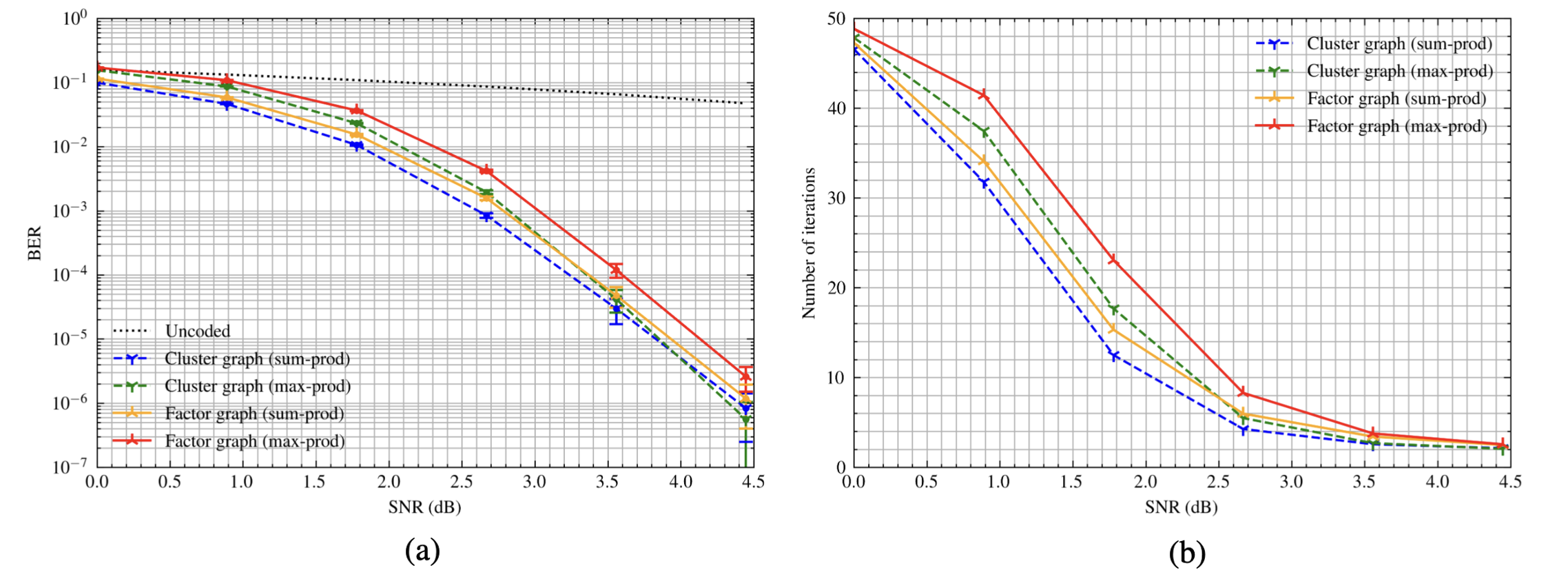}
\caption{Results showing the performance comparison between a factor graph and cluster graph representation of an irregular (220,110) LDPC code.}
\label{fig:results}  
\end{figure*}

The BER comparison between a factor and cluster graph is shown in Fig.~\ref{fig:results}(a). The cluster graph outperforms the factor graph over the entire SNR spectrum with a more pronounced difference at higher SNRs. The cluster graph also outperforms the factor graph when comparing the average number of message passing iterations required by the decoder shown in (b). For both MP and SP inference, cluster graphs outperform their factor graph counterparts. The performance difference between the cluster graph and factor graph is more notable for MP, which can be attributed to the combined effect of slower convergence of factor graphs and slower convergence of MP inference.

For MP, the cluster graph achieves a BER of $10^{-3}$ at 2.8 dB, while the factor graph achieves the same BER at 3 dB. This suggests a coding gain of 0.2 dB for MP inference. 
Similarly, for SP, the cluster graph achieves a BER of $10^{-3}$ at 2.6 dB, while the factor graph achieves the same BER at 2.8 dB. This suggests a coding gain of 0.2 dB for SP inference. 

We also note that both the cluster graph, as well as the factor graph, are stable across the entire SNR spectrum, exhibiting the standard waterfall behaviour (with no error floor region).

Note that the BER performance of SP inference is better than MP inference. This is due to the limit imposed on the maximum number of iterations.

\section{Conclusion and future work}
\label{sec:conclusion}
We proposed a cluster graph representation of LDPC codes using the LTRIP algorithm to improve decoding performance. The cluster graph produces more exact bit marginals compared to the factor graph, which delivers better BER performance. The cluster graph also converges faster. This is achievable without grouping variable or factor nodes to explicitly remove cycles from the LDPC code's graph.

Further research is required to simplify the methods introduced in order to consider them for hardware implementation, which was not in our scope of work. In addition, our message passing schedule may not be the most effective message ordering for cluster graphs in general or its application to LDPC codes. What the most effective message passing schedule is remains open and is a non-trivial question.
\bibliography{IEEEabrv, bibliography}%

\begin{thebibliography}{10}
\providecommand{\url}[1]{#1}
\csname url@samestyle\endcsname
\providecommand{\newblock}{\relax}
\providecommand{\bibinfo}[2]{#2}
\providecommand{\BIBentrySTDinterwordspacing}{\spaceskip=0pt\relax}
\providecommand{\BIBentryALTinterwordstretchfactor}{4}
\providecommand{\BIBentryALTinterwordspacing}{\spaceskip=\fontdimen2\font plus
\BIBentryALTinterwordstretchfactor\fontdimen3\font minus
  \fontdimen4\font\relax}
\providecommand{\BIBforeignlanguage}[2]{{%
\expandafter\ifx\csname l@#1\endcsname\relax
\typeout{** WARNING: IEEEtran.bst: No hyphenation pattern has been}%
\typeout{** loaded for the language `#1'. Using the pattern for}%
\typeout{** the default language instead.}%
\else
\language=\csname l@#1\endcsname
\fi
#2}}
\providecommand{\BIBdecl}{\relax}
\BIBdecl

\bibitem{koller2009probabilistic}
D.~Koller and N.~Friedman, \emph{Probabilistic graphical models: principles and
  techniques}, {F}irst~ed.\hskip 1em plus 0.5em minus 0.4em\relax London W1A
  6US, UK: MIT press, 2009.

\bibitem{kschischang2001factor}
F.~R. Kschischang, B.~J. Frey, and H.-A. Loeliger, ``Factor graphs and the
  sum-product algorithm,'' \emph{IEEE Transactions on information theory},
  vol.~47, no.~2, pp. 498--519, 2001.

\bibitem{colavolpe2022multiuser}
G.~Colavolpe, T.~Foggi, A.~Piemontese, A.~Ugolini, L.~Liu, and J.~Han,
  ``Multiuser detection for time-frequency-packed systems,'' \emph{IEEE
  Transactions on Communications}, vol.~70, no.~10, pp. 6693--6703, 2022.

\bibitem{colavolpe2005application}
G.~Colavolpe and G.~Germi, ``On the application of factor graphs and the
  sum-product algorithm to isi channels,'' \emph{IEEE Transactions on
  Communications}, vol.~53, no.~5, pp. 818--825, 2005.

\bibitem{colavolpe2006ldpc}
G.~Colavolpe, ``On ldpc codes over channels with memory,'' \emph{IEEE
  transactions on wireless communications}, vol.~5, no.~7, pp. 1757--1766,
  2006.

\bibitem{streicher2017graph}
S.~Streicher and J.~{Du Preez}, ``Graph coloring: comparing cluster graphs to
  factor graphs,'' \emph{Proceedings of the ACM Multimedia 2017 Workshop on
  South African Academic Participation}, pp. 35--42, 2017.

\bibitem{streicher2021strengthening}
S.~Streicher and J.~A. du~Preez, ``Strengthening probabilistic graphical
  models: The purge-and-merge algorithm,'' \emph{IEEE Access}, vol.~9, pp.
  149\,423--149\,432, 2021.

\bibitem{lauritzen1988local}
S.~L. Lauritzen and D.~J. Spiegelhalter, ``Local computations with
  probabilities on graphical structures and their application to expert
  systems,'' \emph{Journal of the Royal Statistical Society: Series B
  (Methodological)}, vol.~50, no.~2, pp. 157--194, 1988.

\bibitem{ohtsuki2007ldpc}
T.~Ohtsuki, ``{LDPC} codes in communications and broadcasting,'' \emph{IEICE
  transactions on communications}, vol.~90, no.~3, pp. 440--453, 2007.

\bibitem{jacojohancluster2}
J.~{Du Toit}, J.~{Du Preez}, and R.~Wolhuter, ``{LDPC} codes: tracking
  non-stationary channel noise using sequential variational bayesian
  estimates,'' \emph{arXiv preprint arXiv:2204.07037}, 2022.

\end{thebibliography}
\bibliographystyle{IEEEtran}

\vfill

\end{document}